\documentclass[10pt,twoside,onecolumn,a4paper]{article}

\usepackage{graphicx}  
\usepackage{dcolumn}   
\usepackage{bm}        
\usepackage{amssymb}   

\begin{document}

\title{Chirped Frequency Transfer: A Tool for Synchronisation and Time Transfer}
\author{S. M. F. Raupach and G. Grosche\\Physikalisch-Technische Bundesanstalt (PTB), Bundesallee 100,\\38116 Braunschweig, Germany\\ \textit{\small{e-mail: sebastian.raupach@ptb.de}}}
 
\maketitle

\begin{abstract}
We propose and demonstrate the phase stabilized transfer of a chirped frequency as a tool for synchronisation and time transfer. Technically this is done by evaluating remote measurements of the transferred, chirped frequency. The gates of the frequency counters, here driven by a 10 MHz oscillation derived from a hydrogen maser, play a role analogous to the 1 pulse per second (PPS) signals usually employed for time transfer. In general, for time transfer the gates consequently have to be related to the external clock. Synchronising observations based on frequency measurements on the other hand only requires a stable oscillator driving the frequency counters.\\In a proof of principle, here we demonstrate the suppression of symmetrical delays, such as the geometrical path delay. We transfer an optical frequency chirped by around 240 kHz/s over a fiber link of around 149 km. We observe an accuracy and simultaneity, as well as a precision (Allan deviation, 18000 s averaging interval) of the transferred frequency of around $2\times10^{-19}$. We apply chirped frequency transfer to remote measurements of the synchronisation between two counters' gate intervals. Here, we find a precision of around 200~ps at an estimated overall uncertainty of around 500~ps. The measurement results agree with those obtained from reference measurements, being well within the uncertainty. In the present setup timing offsets up to 4 min can be measured unambiguously. We indicate how this range can be extended further.
\end{abstract}

\section{Introduction}
\begin{figure*}[htbp]
\centering\includegraphics[width=15cm]{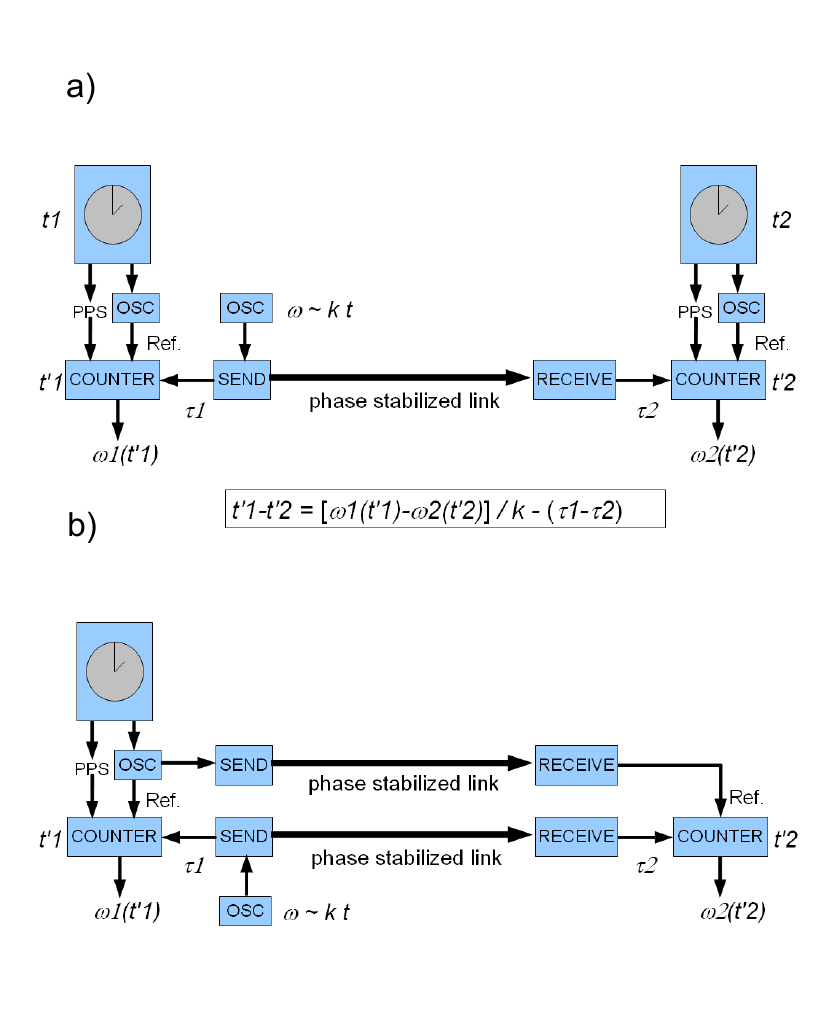}
\caption{Schematic sketch of time transfer using phase stabilized chirped frequency transfer. Key ingredients are a chirp of the frequency, here $\omega = \omega_0 + k t$, and its phase-stabilized transfer realizing a ``zero delay'' link as demonstrated in the proof-of-principle. Panel a) illustrates a comparison between two remote timescales $t'1, t'2$ derived from $t1,t2$. Panel b) illustrates the active transfer of timescale $t'1$, derived from $t1$, to the remote location; OSC: oscillator; PPS: 1 pulse per second signal; Ref.: reference frequency for the frequency counter. } 
\label{fig:generalscheme:activepassive}
\end{figure*}

Correlating independent observations of the same event is at the heart of the scientific quest for revealing the laws of nature. Investigations of how local timescales can be connected in space consequently are pivotal for advancing science. This is the scientific field of time transfer.\\Time transfer can be established using different transmission channels. Transfer via satellites \cite{bauch2006,piester2008}, which typically involves microwave carriers, currently is the only means of achieving global time transfer. On the other hand earth bound optical synchronisation and time transfer via silica fiber links recently has met increased interest, benefiting from high frequency resolution and a low instability \cite{kim2008,whiterabbit, rost2012,lopez2012time,wang2012,sliwczynski2013}.\\At present, time transfer is performed using mainly three approaches \cite{levine2008}: one-way time transfer, two-way time transfer and common-view time transfer. In one-way transfer, a timing signal derived from a local timescale is sent from a local station to a remote one, where it is used for a comparison to the remote timescale; here the path delay experienced by the time signal is assumed to be known with sufficient accuracy by some other means. In two-way transfer, both stations exchange timing signals derived from their respective time scales. The signals are constructed such that timestamps for the reception and transmission of the time signals can be derived for both timescales. From this, the path delay is derived as half of the round-trip travel time \cite{rost2012,levine2008}. In the common-view approach, a ``transfer clock'' is observed by both stations. Adding the transfer of a high-frequency signal and phase measurements \cite{whiterabbit,sliwczynski2013,levine2008} typically allows for a higher resolution. Evaluating the phase of an oscillation at a stable frequency of $1/T$ for time transfer requires some ``pre-synchronisation'' to better than $T$ for avoiding ambiguity. A more detailed discussion of the different approaches can be found in \cite{levine2008}.\\Here we would like to propose and demonstrate a further approach, which we hope to be a useful addition to the toolbox of time transfer. It is based on the actively stabilized transfer of a linearly chirped frequency, counted at both ends of the link. This is illustrated schematically in fig. \ref{fig:generalscheme:activepassive}. The key ingredients are the transmission of a linearly chirped frequency and an active phase stabilization of the link to suppress the effect of the path delay.\\The frequency counters used at the ends of the transmission path define their gate times from a reference oscillator connected to the counters. Here we use a 10 MHz oscillation derived from a hydrogen maser. These gates here are used in analogy to the 1 PPS signals tyically used in metrological time transfer. This grid of internal gates can be related to a grid of 1 PPS signals derived from a clock. Measuring the synchronisation via chirped frequency transfer provides a means for time transfer independent from e.g. the global positioning system (GPS) and can be used to check the results of satellite time transfer.\\Here, we remotely measure the synchronisation of remote counters' gate intervals without using an external time reference. The results are directly applicable to the remote synchronisation of frequency measurements, e.g. for link characterization \cite{predehl2012}, synchronous sampling \cite{takamoto2011,hinkley2013} or \textit{a posteriori} correlation of frequency noise \cite{calosso2013}.\\
The approach demonstrated here does not require a measurement and calibration of the symmetrical path delay \cite{sliwczynski2013} and can be integrated easily into standard setups for optical frequency transfer. Also, it does not require a specialized correlator to exploit pseudo-random codes  \cite{rost2012,lopez2012time} or intricate chirps \cite{boumard2009}. The timing offset is obtained directly from the difference of the frequencies measured at both ends of the links. Using techniques for frequency extraction \cite{grosche2013,bercy2014}, its application can be generalized to network configurations. Its range of unambiguousness can be made ``arbitrarily'' large. It allows for a straighforward definition of simultaneity of remote locations of variable frequency sources, and is well adapted to synchronisation in frequency transfer. Though not demonstrated here, it is expected to intrinsically address the situation of synchronisation measurements between different inertial frames.

\section{Chirped frequency transfer}
\begin{figure*}[htbp]
\centering\includegraphics[width=15cm]{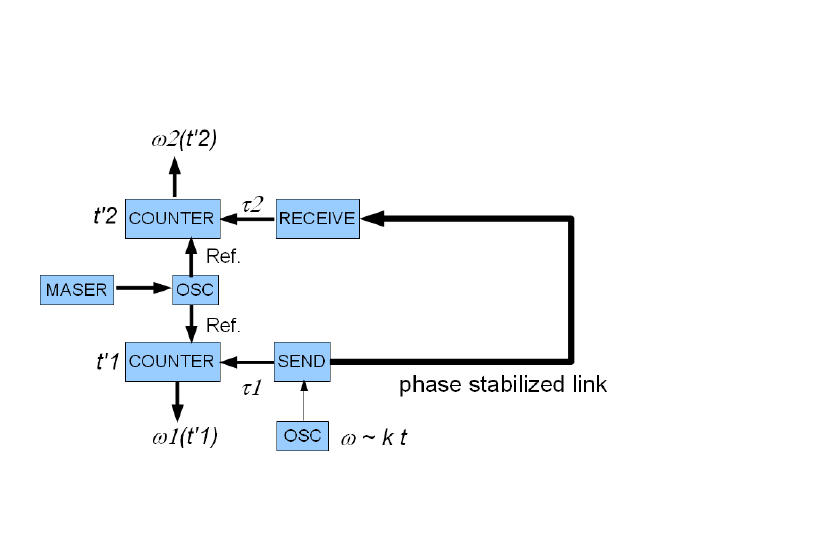}
\caption{Schematic sketch of the loop setup realized in the proof-of-principle experiment. The counters are co-localized and connected to the same oscillator to facilitate verification of the results. The length of the loop is about 149 km.}
\label{fig:generalscheme:loop}
\end{figure*}

Chirped frequency transfer relies on evaluating supposedly ``simultaneous'' measurements of a chirped frequency by two observers, where this frequency is transmitted by the local observer to the remote observer via a phase stabilized link. Here we assume that the timescales of the local end (sender) and the remote end (receiver), $t'1,t'2$ in fig. \ref{fig:generalscheme:activepassive}, are realized using the gates of the local and remote frequency counter.\\From the difference $\Delta\omega$ of the beat frequencies measured at the local end and at the remote end, and from the constant frequency slope $k=\dot\omega$ at the local end, the timescale offset $\Delta t$ is determined as:
\begin{equation}
\Delta t = \frac{\Delta\omega}{k}.
\label{eq:deltaT}
\end{equation}
In eq. \ref{eq:deltaT} we assume, that the time at both ends ``flows at the same speed''. This can be realized e.g. by the stabilized transfer of a constant frequency from the local end to be used for driving the remote frequency counter.\\For achieving a more general form of eq. \ref{eq:deltaT}, we would allow for a time varying scaling factor $\gamma(t)$ of the timescale at the remote end with respect to that at the local end. In this case, eq. \ref{eq:deltaT} would become:
\begin{equation}
\Delta t(t) = \frac{\Delta\omega(t)}{k} = (1-\gamma(t))t+\Delta t.
\label{eq:deltaT:general}
\end{equation}
The most simple example would be oscillators running at slightly different frequencies being used to drive the local and remote frequency counters, thus leading to a constant scaling factor. In these cases, the remote timescale would need to be corrected for $\gamma$, derived e.g. from the slope of the observed frequency difference $\Delta\omega(t)$. Synchronisation is then obtained by shifting the remote timescale such that $\Delta t = 0$. Consequently simultaneity of the measurements at remote locations means $|\omega_1(t'_1) - \omega_2(t'_2)| = \mathrm{const.} = 0$ for $t'_1 = t'_2= t$ and $\gamma(t) = \gamma_0 = 0$.\\
In the proof-of-principle experiment presented here, we realize the case of eq. \ref{eq:deltaT} by performing a loop experiment and using co-localized counters driven by the same oscillator, see fig. \ref{fig:generalscheme:loop}. As will be demonstrated below, active phase stabilization \cite{williams2008} intrinsically adapts to and suppresses symmetrical effects, such as those related to the path length delay along the link. It is interesting to note, that this also applies to synchronisation measurements between different inertial systems, i.e. systems moving at some relative velocity: The backreflected signal reflected off the mirror located in the remote inertial system experiences twice the Doppler shift seen by the remote observer. Therefore the Doppler effect caused by a relative motion at a constant velocity plays the same role as the changing path delay. The Doppler effect adds to it, and hence is expected to being corrected for accordingly by the stabilization. This directly corresponds to the situation of phase-stabilized frequency transfer in a free-space ground to satellite link \cite{djerroud2010,chiodo2013}, which could be extended to time transfer.\\In this paper, however, we restrict ourselves to a practically constant path delay. In the sense that effects of symmetrical delays are intrinsically suppressed, we achieve a ``Zero delay-effect''-link.\\
In the following we will describe the proof-of-principle experiment, where we transfer a linearly chirped optical frequency via an underground fiber link of around 149 km. For the interested reader, we will first technically describe the setup of the optical proof-of-principle experiment in detail. We then demonstrate the transfer of the chirped frequency. In a third step we will demonstrate the remote measurement of the synchronisation between the local and the remote timescale, as represented by the gates of the respective frequency counters.

\subsection{Experimental Setup}
\label{Setup}
\begin{figure*}[htbp]
\centering\includegraphics[width=8.8cm]{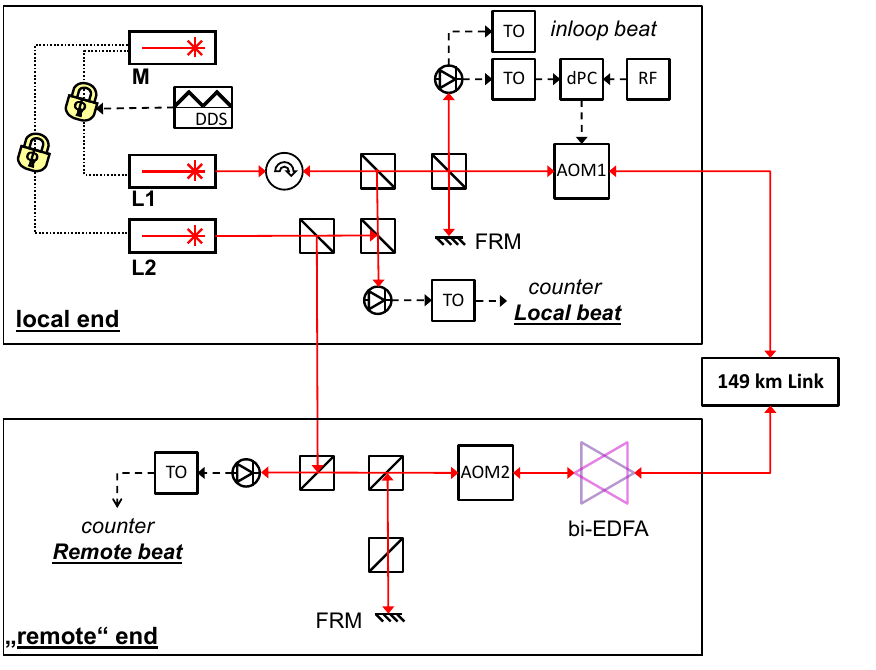}
\caption{Schematic sketch of the optical, all-fiber setup for phase-stabilized transfer of a chirped optical frequency; TO: tracking oscillator; DDS: direct digital synthesizer; dPC: digital phase comparator; RF: reference radio frequency source; AOM: acousto-optical modulator (AOM1: +55 MHz,  AOM2: +40 MHz); FRM: Faraday rotator mirror; bi-EDFA: bidirectional Erbium-doped amplifier; M,L1,L2: lasers; beamsplitter symbols represent fiber couplers; counter channels for DDS output and the double-tracked inloop beat frequency not depicted. The counters are driven by a radio frequency of 10 MHz derived from the same oscillator.}
\label{fig:scheme}
\end{figure*}

The optical setup is shown schematically in figure \ref{fig:scheme}. To facilitate verification of the results, a pair of single mode, standard telecom silica fibers connecting PTB (Braunschweig/Germany) and Leibniz University (Hanover/Germany) are patched to form a loop. Both ends of the loop are located in the same laboratory at PTB. Two infrared fiber lasers (L1, L2; around 194.4 THz / 1542 nm) are phase locked to the same master laser (M). This allows full control over their relative frequency. In the lock of L1, the reference radio frequency (rf) delivered by a direct digital synthesizer (DDS; AD9956, 400 MHz clock, 48 bit) can be tuned (step size here around 23.84 Hz). Thereby the optical frequency of L1 is tuned relative to that of L2. The applied frequency tuning here is about $\pm$238 418.6 Hz/s over a range of 20 MHz. Here, in between the chirps there is also an interval of about 30 s, where the frequency remains constant (see fig. \ref{fig:unstabilized:stabilized}, blue curve).\\
For stabilization of the fiber link \cite{williams2008}, the light enters a fiber Michelson interferometer. It is formed by the 149~km fiber link, and a second, short fiber arm serving as a reference for the phase of the light travelling along the link and back. For correcting for the measured deviations, an acousto-optic modulator is used at the input of the fiber link. Care is taken to place the photodetectors for the local and the remote beat at about the same distance with respect to the respective, local Faraday rotator mirror to within 0.02 m. The total systematic uncertainty of the delay due to path length differences is estimated conservatively as 250 ps. The differential delay between the photodetectors is measured by simultaneously detecting the same beatnote on both detectors. It amounts to $(450\pm100)$~ps.\\Unless e.g. an optical clock is located at both ends of the link, also the constant frequency of L2 would have to be transferred to the remote end \cite{predehl2012,williams2008,lopez2012frequency}. At the remote end, the beat frequency $\omega2(t'2)=|\nu_{\mathrm{L1}}-\nu_{\mathrm{L2}}|$ could be detected either directly on a photodetector, or a virtual beat frequency could be obtained using heterodyne detection and an optical transfer oscillator \cite{telle2002}. The transfer of the frequency could be done either on the second fiber of the pair or, taking advantage of common mode suppression of fiber noise, in the same fiber. This can be implemented by using radio frequency transfer \cite{lopez2010} or single-sideband \cite{schediwy2012} or two-carrier transfer using e.g. appropriate optical \cite{rohde2013,zou2013} or rf filters for the inloop beat notes. In these implementations, the residual unsuppressed fiber noise of the long link would be common mode to both transferred frequencies. Also, at the remote end $\nu_{\mathrm{L2}}$ could be used further for the remote generation of a local ``timetick'' by deriving an ultrastable radio frequency from it using a frequency comb \cite{fortier2011}.\\In this proof-of-principle experiment only the varying frequency of L1 is transferred over the link. Separate, short pieces of unstabilized fiber guide the light of L2 to the photodetectors, see fig. \ref{fig:scheme}. Here, in the absence of common mode suppression the residual unsuppressed fiber noise of the 149 km link fully enters the result.\\Two separate, dead-time free frequency counters operated in $\Lambda$-mode \cite{kramer2001,dawkins2007} are used to measure the local and remote beat frequencies. The counters are connected to the same reference source (10 MHz, derived from a hydrogen maser). We change their timing offset by switching them on at arbitrary points in time and by changing the length of the cable delivering the 10 MHz driving frequency.\\In the following we first discuss the results of high fidelity transfer of the chirped frequency, before presenting the results from synchronisation measurements. 

\subsection{Simultaneity of the Chirped Frequency}
\label{sec:results}
\begin{figure*}[htbp]
\centering\includegraphics[width=8.8cm]{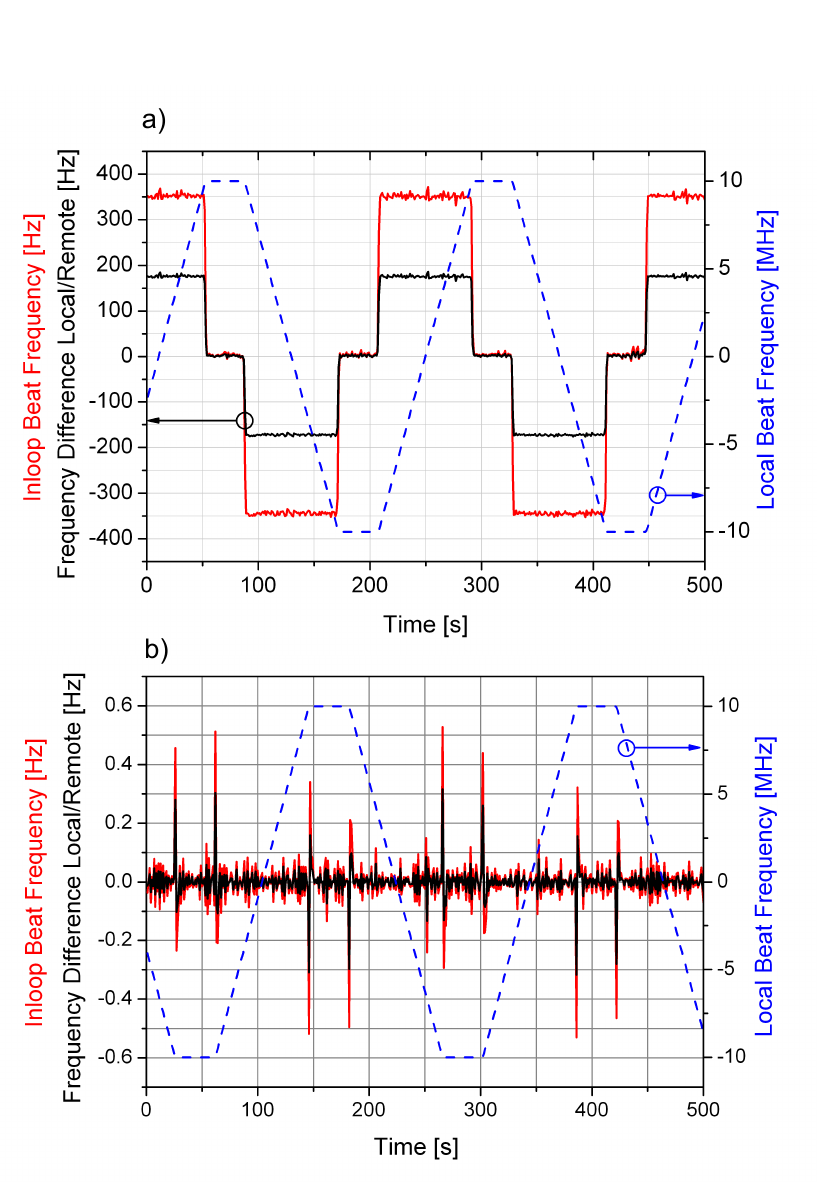}
\caption{Inactive and active link stabilization. Panel a) illustrates the unstabilized case. It shows the deviation of the inloop beat frequency (red) from its center value as well as that of the difference between the local and remote beat frequency (black) and of the local beat frequency (blue) without the stabilization loop being active. Panel b) illustrates the case of active link stabilization. Shown are again the difference between the local and remote beat frequency (black) and the deviations of the local beat frequency (blue) and of the inloop beat frequency (red) from their respective center frequency (here: 47.5 MHz, and 29 MHz). The local beat frequency shows how L1 is tuned relative to L2. The repetition interval is around 4~min. The local and remote beat frequencies here are measured on two channels of the \emph{same} counter. The difference between local and remote beat frequency illustrates that this chirp is transferred faithfully by the active stabilization, represented by the inloop beat frequency.}
\label{fig:unstabilized:stabilized}
\end{figure*}

\begin{figure*}[htbp]
\centering\includegraphics[width=8.8cm]{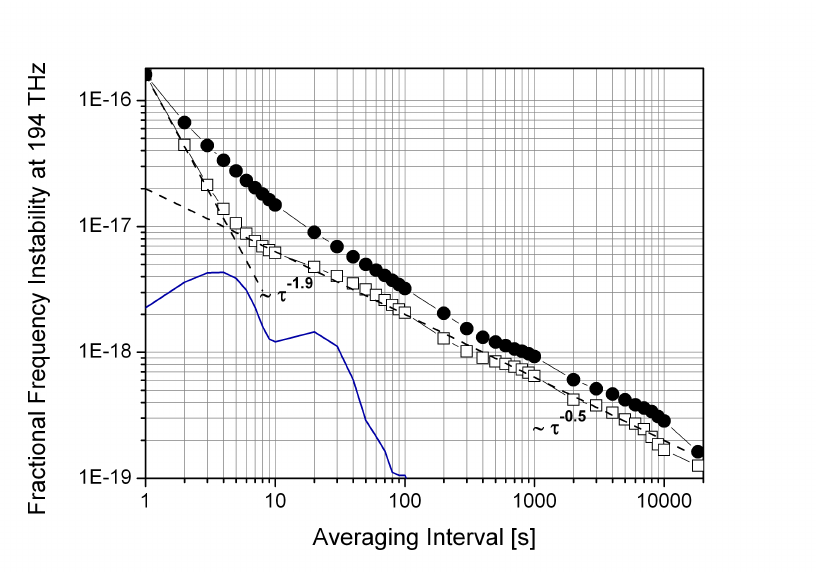}
\caption{Fractional instability of the chirped frequency transfer: modified Allan deviation (open squares), and (overlapping) Allan deviation for unweighted averaging (full circles). The blue line indicates the modified Allan deviation of the estimated differential instability contribution of the local and remote tracking oscillators. The dashed lines are guides to the eye. The frequency data here are measured on different channels of the same $\Lambda$-type frequency counter with a gate time of 1 s.}
\label{fig:ADEV:samecounter}
\end{figure*}

The simultaneity of the local and remote frequencies for stabilized frequency transfer is illustrated in fig. \ref{fig:unstabilized:stabilized} as discussed in the following.\\For the unstabilized case, see panel a), we observe an inloop beat frequency offset of $\pm(348.8\pm1.3)$ Hz, where the uncertainty is given by the instability (modified Allan deviation) of the frequency values. From the frequency slope of around 238~418.6~Hz/s, the inloop beat frequency offset corresponds to a signal delay of around $1.463\pm0.006$~ms. Using a group index of refraction of 1.468 as specified for SMF28 fibre, this yields a length of the round-trip of $(298.8\pm1.1)$~km. This agrees with a one-way length of around 148.8 km, obtained from optical time domain reflectometry.\\
For the stabilized case, the absence of such a delay related frequency offset is illustrated by the inloop beat frequency shown in panel b) of figure \ref{fig:unstabilized:stabilized} (red curve; note the difference in scales).\\
For data analysis, only the central $\pm8$ MHz of each chirp (blue curve) are used, corresponding to a chirp time of 67 s. No further data are rejected. Note that from eq. \ref{eq:deltaT}, when averaging over frequency chirps of opposite sign, a constant time delay would cancel out when calculating the average of the frequency difference offset $\Delta \omega$, while for the calculation of $\Delta t$ a constant frequency offset cancels out.\\In a first step, the remote and the local beat frequencies both are measured simultaneously on the \emph{same} counter (figure \ref{fig:unstabilized:stabilized}, black curve). This allows determining the instability and accuracy of the frequency transfer, i.e. the simultaneity of the frequencies at both ends. We calculate the unweighted mean and the Allan deviation of the frequency difference offset, as well as the modified Allan deviation of $\Delta\omega$ \cite{dawkins2007,riley2008}, shown in figure \ref{fig:ADEV:samecounter}. 
The modified Allan deviation (modADEV) initially averages down slightly faster than for pure white phase noise \cite{droste2013}, and is $\propto1/\sqrt{\tau}$ for large $\tau$, because of averaging over separate frequency intervals. The unweighted average value of the fractional frequency difference is $1.92\times 10^{-19}$ (around 73000 data points of 1~s averaging time each, not including the ``dead time'' data points in between the analysed parts of the chirps). While this compares well with results from long-distance stable frequency transfer \cite{predehl2012,droste2013}, this value is slightly larger than that of the last point on the instability curve (Allan Deviation at an interval of 18000 s), which is $1.62\times 10^{-19}$ (modADEV: $1.25\times 10^{-19}$), and may hint at a systematic offset on the $10^{-19}$-level. Individually, the mean fractional frequency values for chirps with opposite sign are $+9.22\times 10^{-19}$ and $-5.38\times 10^{-19}$. These offsets indicate the presence of undetected, chirp-dependent effects. In particular delays introduced in the process of detection of the local and remote beat signal are not detected and are not suppressed by the stabilization. We have identified the tracking oscillators as a significant and variable source of delay not related to the link, and their differential delay is calibrated at each measurement run.

\subsection{Remote Synchronisation Measurement}
\begin{figure*}[htbp]
\centering\includegraphics[width=8.8cm]{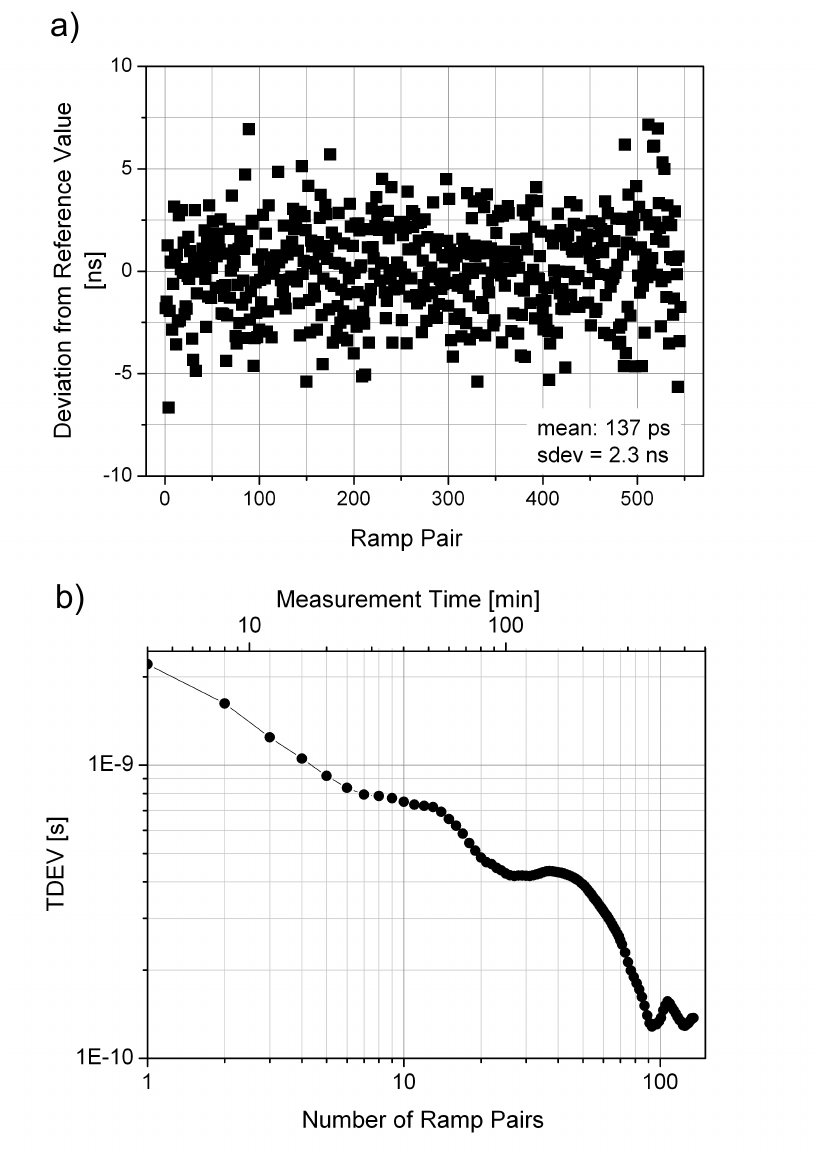}
\caption{Measuring the time offset via the fiber link. Panel a) shows the deviations of the unweighted averages over pairs of consecutive chirps having opposite sign, i.e. an analysed chirp time per point of around 134~s, and a total measurement interval per point of around 4~min (see text); panel b) indicates the according instability (time deviation (TDEV)) of the data shown in panel a). Note that here residual fiber noise fully enters the result, as only the chirped frequency is transferred via the link.}
\label{fig:timedifference:localremote}
\end{figure*}

\begin{figure*}[htbp]
\centering\includegraphics[width=8.8cm]{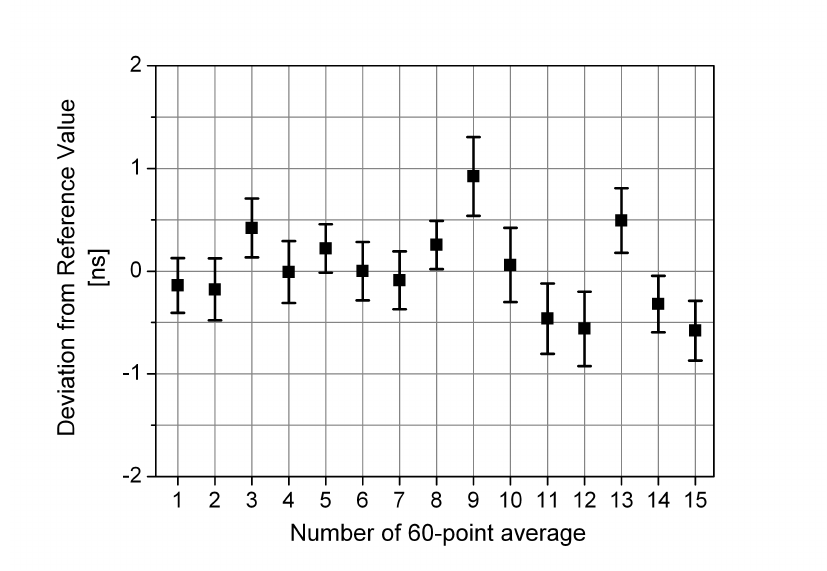}
\caption{Scatter of the combined data of three measurement runs. Shown are the unweighted averages over 60 consecutive pairs of chirps, corresponding to a measurement time of 4 hours per data point. Error bars indicate the standard error (standard deviation of the mean) for each point. See text for details.}
\label{fig:subintervals}
\end{figure*}

For demonstrating the remote measurement of timescale synchronisation, the remote and local beat frequencies are measured on \emph{different} counters.\\
Between the runs we change the offset between the frequency counters' timescales by briefly interrupting the reference of one of the counters, and by changing the cable length. Also, we swap the counters between measurement run 1 and 2. We perform three separate measurement runs, lasting for around 36.5 hours, 11 hours and 12.5 hours, respectively.\\To assess the accuracy of the synchronisation measurement, for each run we obtain reference values by directly connecting both counters to one tracking oscillator tracking the chirped output of the DDS. We determine the reference value for the offset of the ``timescales'', represented by the gates of the frequency counters, from this direct side-by-side measurement. This is done by performing the chirped frequency analysis according to eq. \ref{eq:deltaT}, i.e. using the same analysis as for the case of optical frequency transfer. For this laboratory measurement, we obtain a statistical uncertainty of around 1~ps and a systematic uncertainty of order 10~ps. This constitutes a method for precise synchronisation of frequency counters over short distances. When measuring the reference values, care is taken to use the same electrical cables as in the measurement via the link.\\Furthermore, the counters offer the feature to directly extract a copy of their internal 1 pulse-per-millisecond (1 pp-ms) grid. This allows obtaining a second, coarse reference value by measuring the 1 pp-ms signals on a digital oscillocope. In the process, we discovered an internal asymmetry of the two counters' internal processing of the 10 MHz reference. This led to a total asymmetry of around 16 ns between the internal 1 pp-ms grid and that measured externally, which is corrected for.\\The corrected data of the first run are shown in figure \ref{fig:timedifference:localremote}. Figure \ref{fig:timedifference:localremote}a shows the scatter of the data for pairs of one positive and one negative chirp. Each data point therefore corresponds to an analysed chirp time of about 134 s, while the total chirp time per ramp pair would be about 3 min and the total measurement interval per ramp pair is about 4 min, including the times of unchirped frequency. Figure \ref{fig:timedifference:localremote}b shows the according time deviation (TDEV) as a measure of statistical uncertainty \cite{rost2012}. The time deviation seems to reach a flicker floor below 200 ps, where the last point of the TDEV is 140 ps. The mean of the corrected data shown in fig. \ref{fig:timedifference:localremote}a is around 140 ps, comparable to the uncertainty given by the instability.\\
 The results of all three runs are listed in table \ref{tab:results}. The corrections applied to the synchronisation measurement via the link are those arising from the calibration of the differential delay introduced by the tracking oscillators and for the differential delay introduced by the remote and inloop photodetectors. The systematic uncertainties include the sensitivity of the results to determining the frequency slope from the data using either the previous or the following data point. For active synchronisation, where $\Delta t$ would be minimized iteratively, this contribution would be minimized accordingly. Another major contribution to the systematic uncertainty arises from the variable delay introduced by the tracking oscillators, leading to an uncertainty contribution of 230 ps. Statistical uncertainties are the instabilities obtained from the time deviation, calculated using commercial software.\\For the 1 pp-ms measurements, we apply corrections for the aforementioned asymmetry and the inaccuracy of the oscilloscope's internal clock, which is calibrated during each measurement.\\The results of the proposed method agree well with those from direct side-by-side measurements, demonstrating the viability of the proposed method.\\Effects which are non-symmetric, such as dispersion asymmetric with respect to the outgoing and the returning path, or which are not cumulative with respect to the roundtrip path (i.e. the sum of outgoing and returning path) cannot be suppressed by the stabilization. The most prominent example is the rotational delay (Harress-Sagnac effect) in a rotating reference frame \cite{footnoteHarressSagnac,laue1911,sagnac1913a,sagnac1913b,harzer1914,knopf1920,laue1919}, such as Earth rotating around its axis \cite{post1967,allan1985}. If required, such effects would have to be corrected for separately in a point-to-point connection. However, as discussed in the appendix, due to the small frequency differences involved in our method, the effect of chromatic dispersion is well below 1 ps here. The overall potential effect of dispersive asymmetries on the time transfer is estimated conservatively to be not more than 10 ps. Also, the rotational delay in our experimental setup is negligible, where the area enclosed by the 149 km parallel fiber loop, and accordingly its projection onto the equatorial plane, is very small.\\To assess the overall statistical uncertainty of the measurements, we combine the corrected data of all three runs, and subdivide the combined set into measurement intervals of around 4 hours each. The result is shown in figure \ref{fig:subintervals}. 
The overall mean is +3 ps with a standard deviation of the mean of 110 ps, while the overall median is -7 ps with a median absolute deviation of 260 ps. Calculating the total time deviation \cite{riley2008} of the data indicates an overall instability of around 210 ps. 

\subsection{Technical Improvements}
The results achieved in this proof-of-principle experiment do not yet achieve the best results obtained in other experiments on optical time transfer via fiber \cite{rost2012,lopez2012time,sliwczynski2013}.\\Using advanced equipment from two-way satellite time and frequency transfer (TWSTFT) \cite{rost2012,lopez2012time}, two-way time transfer was demonstrated, along with optical stable-frequency transfer \cite{lopez2012time}, over up to 540 km of installed fiber. For this distance statistical uncertainties of around 10 ps (time deviation, (TDEV) \cite{riley2008}) and a total uncertainty of 250~ps were reported. In a second approach \cite{sliwczynski2013}, the delay of timing signals sent over a fiber link is calibrated and actively kept constant. For advanced laboratory measurements using up to 480 km of fiber, record statistical uncertainties (TDEV) down to 0.5 ps \cite{sliwczynski2013} were reported with a total uncertainty of around 23~ps. For data network synchronisation, also techniques for simultaneous transfer of network traffic and time are being investigated \cite{sotiropoulos2013}.\\
The proof-of-principle setup used here could be improved in several ways to achieve higher performance.\\Here only the chirped frequency was actually transferred via the fiber link, such that residual unsuppressed fiber noise fully enters the ``remote'' beat note between the stable and the chirped frequency, see section \ref{Setup}. In that sense the present setup realizes a worst case. Transferring both frequencies in the \emph{same} fiber will suppress the relative frequency noise improving the precision and reducing the required averaging time. This is similar to the stabilized transfer of a single sideband modulated optical carrier. For the transfer of a microwave frequency amplitude-modulated onto an optical carrier, an absolute precision of around $10^{-8}$ Hz was reported \cite{lopez2010}. If this precision can be maintained also for \emph{chirped} microwave frequency transfer via optical fiber, according to eq. \ref{eq:deltaT} for a chirp of 100~kHz/s this would allow for sub-picosecond precision of chirped-frequency time transfer.\\We note that taking advantage not only of the stabilized frequency but also of the stabilized phase would immediately increase the attainable precision further.
Also, longer chirps could be used to increase the duration of continuous measurement, i.e. to achieve a faster averaging of the instability, but would require an accordingly large acceptance range of the frequency counters. Furthermore, for a given absolute frequency precision (in [Hz]), the resolution scales with the slope of the frequency chirp, see eq. \ref{eq:deltaT}. Therefore, using a stronger chirp \cite{chiodo2013} would be advantageous. Note, that for a constant delay the correction applied to the outgoing frequency by the stabilization loop is \emph{not} a chirp. It is also a - apart from the fiber noise - constant offset corresponding to the inloop frequency offset shown in panel a) of fig. \ref{fig:unstabilized:stabilized}.\\To reduce the systematic uncertainty, the detection electronics as well as the tracking oscillators would have to be replaced or developed further (here standard equipment from ultrastable frequency transfer was used). In fact, when approaching sub-100~ps accuracy, the calibration of the local equipment as well as the stability of this calibration will become highly demanding.

\section{Conclusions}
We have proposed and in a proof-of-principle experiment demonstrated an approach to simultaneous transfer of time and optical frequencies. It is based on the phase-stabilized transfer of a chirped frequency.\\
In the proof-of-principle experiment we transferred a chirped optical continuous-wave frequency via a fiber link of around 149 km. The frequency was chirped linearly by about 240 kHz/s. We observed the suppression of symmetrical delays. Using two channels on the same frequency counter, we found an accuracy, indicating simultaneity, and a precision of chirped frequency transfer of around $2\times10^{-19}$. In a second step we employed chirped frequency transfer for remote timescale synchronisation measurements. We here found a precision of around 200 ps, and an estimated overall uncertainty of around 500~ps. The results agreed with reference measurements, and were well within the estimated overall uncertainty. We have outlined straightforward technical improvements.\\Compared to other approaches to time transfer, the demonstrated approach does not require a correlator or a calibration of symmetrical delays. It is particularly well-adapted to the synchronisation of frequency measurements, as the gate intervals of the frequency counters directly can be treated as the realizations of the timescales. In the present setup the time interval within which the synchronisation can be measured unambiguously is around 4 min. This range is limited by the repetition intervals of the chirps. However, in principle the demonstrated technique does not require a periodic modulation. Instead, a single, long chirp would be sufficient, allowing for unambiguous synchronisation within ``arbitrarily'' long time intervals, constrained by the duration of the chirp only.

\section*{Acknowledgements}
We acknowledge helpful comments by U. Sterr and A. Bauch, and thank A. Koczwara and M. Misera for excellent technical support. Financial support by the European Metrology Research Programme (SIB-02, ``NEAT-FT'') and the Centre for Quantum Engineering and Space-Time Research QUEST is gratefully acknowledged. The EMRP is jointly funded by the EMRP participating countries within EURAMET and the European Union.

\section*{Appendix}
Here we discuss sources of asymmetric delays experienced by the light traveling along the link, that may affect phase-stabilized chirped-frequency transmission. In the following we assume that the optical frequency $\nu$ is chirped linearly.\\Before the feedback loop for link stabilization is active, i.e. when $\nu_{\mathrm{input}}(t)\neq \nu_{\mathrm{remote}}(t) \neq \nu_{\mathrm{inloop}}(t)$, the frequencies before and after a roundtrip, as well as the remote frequency, at time $t$ are given by:
\begin{eqnarray}
\label{eq:InputRemoteInloop}
\nu_{\mathrm{input}}(t)&=& \nu_0 + k ~t\\
\nu_{\mathrm{remote}}(t)&=& \nu_0 + k ~(t - \tau_{\mathrm{Link,remote}})\\
\nu_{\mathrm{inloop}}(t)&=& \nu_0 + k ~(t - \tau_{\mathrm{Link,roundtrip}})
\end{eqnarray}
Here, $k$ is the slope of the frequency chirp, $\tau_{\mathrm{Link,remote}}$ is the delay caused by the propagation along the link to the remote end in the fiber, and $\tau_{\mathrm{Link,roundtrip}}$ is the delay corresponding to a roundtrip of the light to the remote end and back to the local end.\\The one-way delay $\tau_{\mathrm{Link,remote}}$ is given by:\\
\begin{eqnarray}
\tau_{\mathrm{Link,remote}} &=& \tau_0\nonumber\\&& + \tau_{\mathrm{CD}}(L,\nu_{\mathrm{remote}}(t))\nonumber\\&&+ \tau_{\mathrm{PMD,outgoing}}\nonumber\\&&+ \tau_{\mathrm{rot}},
\end{eqnarray}
where
\begin{equation}
\label{eq:tau0}
\tau_0 = \frac{n(\nu_0)L}{c}.
\end{equation}
Here $c$ is the vacuum speed of light, $n(\nu_0)$ is the group index of refraction, see e.g. section \ref{sec:results}, at $\nu_0$; $L$ is the geometric length of the fiber, $\tau_{\mathrm{PMD}}$ is the polarization-dependent delay caused by birefringence of the fiber, and $\tau_{\mathrm{rot}}$ is the rotational delay (Harress-Sagnac effect) \cite{sagnac1913a,sagnac1913b,harzer1914,knopf1920,post1967,allan1985}; $\tau_{\mathrm{CD}}(\nu_{\mathrm{remote}}(t))$ is the differential delay experienced by the light at $\nu_{\mathrm{remote}}(t)$ relative to light at $\nu_0 $ due to chromatic dispersion, where $\nu_{\mathrm{remote}}(t)=\nu_{\mathrm{input}}(t-\tau_{\mathrm{Link,remote}})$. In the following we assume linear dispersion, where $D_{\mathrm{CD}}$ (in units of [ps/(nm km)]) is the group velocity dispersion constant for a difference in wavelength $\Delta\lambda$ and a fiber length $L$.\\
Hence:
\begin{eqnarray}
\label{eq:tCD:remote}
\tau_{\mathrm{CD}}^{\mathrm{remote}} &=& D_{\mathrm{CD}}~L~(\lambda_{\mathrm{remote}} - \lambda_0)\nonumber\\
                    &=& D_{\mathrm{CD}}~L~c\left(\frac{1}{\nu_{\mathrm{remote}}(t)} - \frac{1}{\nu_0}\right)\nonumber\\
                    &\approx& D_{\mathrm{CD}}~L~c \frac{\nu_0 - \nu_{\mathrm{remote}}(t)}{\nu_0^2}\nonumber\\
                    &=& -D_{\mathrm{CD}}~L~c \frac{D_0~(t - \tau_{\mathrm{Link,remote}})}{\nu_0^2}\nonumber\\
                    &\approx& -D_{\mathrm{CD}}~L~c~D_0 \frac{t - \tau_0}{\nu_0^2}.
\end{eqnarray}
The contribution of chromatic dispersion to the frequency measured at the remote end is: 
\begin{equation}
|\Delta\nu_{\mathrm{CD}}| \approx  \left|-D_{\mathrm{CD}}~L~c~D_0^2 \frac{t - \tau_0}{\nu_0^2}\right|.
\end{equation}
In the last line of eq. \ref{eq:tCD:remote} we have used that typically $\tau_{\mathrm{CD}}, \tau_{\mathrm{PMD}} \ll \tau_0$. Furthermore, we used that even for a east-/westward fiber link, yielding the maximum rotational delay, we would have:
\begin{equation}
\frac{\tau_{\mathrm{rot}}}{\tau_0} = \frac{R_E\cos(\phi)\Omega}{nc},
\end{equation}
where $R_E$ is Earth's radius, $\phi$ is the latitude, $n$ is the refractive index, and $\Omega$ is Earth's angular velocity. As an example, for a latitude of 52$^\circ$ (Braunschweig) and using $R_E \approx 6371$ km, and $\Omega \approx 7.29\times10^{-5}$~/s,  this yields $\tau_{\mathrm{rot}}/\tau_0~\approx7\times10^{-7}$. Even on the equator this ratio would be around $10^{-6}$ only, so here we can safely neglect $\tau_{\mathrm{rot}}$ relative to $\tau_0$.\\
The delay caused by polarization mode dispersion will be discussed below.\\
The delay of the roundtrip signal can be written as:
\begin{eqnarray}
\tau_{\mathrm{Link,inloop}} &=& 2\tau_0\nonumber\\
&& + \tau_{\mathrm{CD}}^{\mathrm{outgoing}}(L,\nu_{\mathrm{inloop}}(t))\nonumber\\
&& + \tau_{\mathrm{rot}}\nonumber\\
&& + \tau_{\mathrm{CD}}^{\mathrm{return}}(L,\nu_{\mathrm{inloop}}(t)+2\nu_{\mathrm{AOM}})\nonumber\\
&& - \tau_{\mathrm{rot}}\nonumber\\
&& + \tau_{\mathrm{PMD,outgoing}} + \tau_{\mathrm{PMD,return}},
\end{eqnarray}
where $\tau_{\mathrm{PMD,outgoing/return}}$ are the delays caused by polarization mode dispersion of the outgoing and returning light, $\nu_{\mathrm{inloop}}(t) = \nu_{\mathrm{input}}(t - \tau_{\mathrm{Link,roundtrip}})$, and $2\nu_{\mathrm{AOM}}$ is the frequency shift induced by double-passing the acousto-optic modulator (AOM) at the remote end. From equation \ref{eq:tCD:remote}:
\begin{equation}
\tau_{\mathrm{CD}}(L,\nu_{\mathrm{inloop}}(t)) \approx -D_{\mathrm{CD}}~L~c \frac{D_0~(t - 2\tau_0)}{\nu_0^2},
\end{equation}
and
\begin{eqnarray}
\tau_{\mathrm{CD}}(L,\nu_{\mathrm{inloop}}(t)+2\nu_{\mathrm{AOM}})\approx&-D_{\mathrm{CD}}~L~c \frac{D_0~(t - 2\tau_0)}{\nu_0^2}\nonumber\\
&-D_{\mathrm{CD}}~L~c\frac{2\nu_{\mathrm{AOM}}}{\nu_0^2},
\end{eqnarray}
where $D_0~(t - 2\tau_0) \approx \nu_{\mathrm{inloop}}(t) -\nu_0$.
Hence
\begin{eqnarray}
\label{eq:tCD:inloop2}
\tau_{\mathrm{CD}}^{\mathrm{inloop}} &=&2~\tau_{\mathrm{CD}}^{\mathrm{remote}}\nonumber\\
                                     &&-D_{\mathrm{CD}}~L~c\frac{2\nu_{\mathrm{AOM}}}{\nu_0^2}\nonumber\\
                                     &&-2~D_{\mathrm{CD}}~L~c\frac{D_0~\tau_0}{\nu_0^2}
\end{eqnarray}
As can be seen from equation \ref{eq:tCD:inloop2}, there are two sources of delay asymmetry due to chromatic dispersion. From eq. \ref{eq:tau0}, the last term can be expressed as $2~D_{\mathrm{CD}}~L^2~n(\nu_0)~D_0/\nu_0^2 \approx 10^{-23}$s. The frequency shift by the AOM on the other hand might introduce a noticeable asymmetry of the delay, half of which consequently would be imposed onto the stabilized signal by the stabilization loop. For a link length of 149 km, an AOM-frequency of 40 MHz and a typical value for the chromatic dispersion of standard single mode fiber (G.652) at 1542 nm of around 16.6 ps/(nm~km) \cite{itug652}, the resulting asymmetry is around 1.6 ps. Consequently the stabilization may be expected to impose a synchronisation mismatch of around 800 fs at the remote end. This can be modelled or measured \cite{sliwczynski2013} and corrected for. Furthermore, if required this contribution could easily be reduced further here by e.g. choosing a smaller offset frequency for the frequency shifter at the remote end.\\A separate source of asymmetry is polarization mode dispersion: As the light is reflected by a Faraday rotator mirror at the remote end, the polarizations of the outgoing and the returning light are orthogonal. Generally the delays might be asymmetric, i.e. $\tau_{\mathrm{PMD,outgoing}} - \tau_{\mathrm{PMD,return}}\neq 0$. Using a link design value of around 0.1 ps/km$^{1/2}$ \cite{sliwczynski2013}, a value of the order of 1 ps would be estimated, while measurements over 540 km of installed fiber \cite{lopez2012time} indicate a somewhat higher uncertainty due to PMD. For a 149 km link this would correspond to around 10 ps. This could be suppressed further by averaging over consecutive measurements performed with orthogonal input polarizations.\\Finally, the rotational delay is conventionally corrected for, i.e. the rotating system is effectively transformed into a non-rotating one. In our experiment the rotational effect due to Earth's rotation is negligible, as both ends of the link are located in the same laboratory, and the fibers are running in parallel in the same bundle, with a negligible net distance difference orthogonal to Earth's rotational axis. For a point-to-point east-/westward link of $L=$150 km at the same latitude, we obtain a $\tau_{\mathrm{rot}}= L R_E\Omega\cos(\phi) / c^2 $ between the sending end and the remote end of around 500~ps. We note that the rotational delay does not depend on the index of refraction \cite{laue1919,wang2003,arditty1981}.\\Finally we note that in case of linear dispersion the slope of both the inloop and the remote frequency time trace remains linear. With respect to non-linear dispersion it is interesting to note that even for a link of around 1000 km (one-way delay $\tau$: 5 ms) and a ramp speed of 2 MHz/s, the relevant frequency range covered within the reciprocal of the feedback loop's bandwidth $4\tau$ \cite{williams2008} is 40 kHz only. Hence the effect of non-linear dispersion will in general be quite small. If using non-linear chirps of the input frequency, the feedback loop used for stabilization of the fiber link may ultimately benefit from a corresponding adaption to the continuously changing slope.

\clearpage
\listoffigures
\clearpage

\newpage
\newpage

\begin{table*}
\caption{Results from and corrections applied to three measurement runs performed within a period of one week. Listed are the corrected results of measuring the time offset via the phase stabilized fiber link, as well as the reference values obtained from direct side-by-side measurements. The statistical (u$_{\mathrm{A}}$) and systematic (u$_{\mathrm{B}}$) uncertainties are those of the corrected values.}
\label{tab:results}
\small
\begin{tabular}{llcccc}
&&measurement result&applied correction&u$_{\mathrm{A}}$&u$_{\mathrm{B}}$\\
\hline
Run 1&Link measurement (corrected)&\bf{- 0.370 509 891 66 s}&-590 ps& 140 ps&390 ps\\
&direct measurement&\bf{- 0.370 509 891 80 s}&-&3 ps&40 ps\\
&1 pp-ms measurement (corrected)&\bf{- x.xxx 509 891 92 s}&+16350 ps&-&720 ps\\
\hline
Run 2&Link measurement (corrected)&\bf{- 0.750 319 001 35 s}&-690 ps&300 ps&505 ps\\
&direct measurement&\bf{- 0.750 319 001 08 s}&-&2 ps&50 ps\\
&1 pp-ms measurement (corrected)&\bf{- x.xxx 319 000 45 s}&-16200 ps&-&720 ps\\
\hline
Run 3&Link measurement (corrected)&\bf{- 0.219 302 701 27 s}&-960 ps&360 ps&370 ps\\
&direct measurement&\bf{- 0.219 302 701 13 s}&-&3 ps&10 ps\\
&1 pp-ms measurement (corrected)&\bf{- x.xxx 302 700 44 s}&-16140 ps&-&720 ps\\
\hline
\end{tabular}
\normalsize
\end{table*}

\begin{table*}
\caption{List of uncertainty contributions and corrections applied to the link measurement results presented in table \ref{tab:results}}
\label{tab:corrections}
\small
\begin{tabular}{llccc}
&&applied correction&u$_{\mathrm{A}}$&u$_{\mathrm{B}}$\\
\hline
Run 1&Measurement \& analysis& 0~ps&140~ps&170~ps\\
&Tracking oscillators&-140~ps&0.7~ps& 230~ps\\
&photo detectors&-450~ps&3~ps&100~ps\\
&fiber paths&0~ps&0~ps&250~ps\\
\hline
Run 2&Measurement \& analysis& 0~ps&300~ps&360~ps\\
&Tracking oscillators&-240~ps&1~ps& 230~ps\\
&photo detectors&-450~ps&3~ps&100~ps\\
&fiber paths&0~ps&0~ps&250~ps\\
\hline
Run 3&Measurement \& analysis& 0~ps&360~ps&150~ps\\
&Tracking oscillators&-510~ps&1~ps& 230~ps\\
&photo detectors&-450~ps&3~ps&10~ps\\
&fiber paths&0~ps&0~ps&250~ps\\
\hline
\end{tabular}
\normalsize
\end{table*}

\begin{thebibliography}{99}
\bibitem{bauch2006} A. Bauch et al., ``Comparison between frequency standards in Europe and the USA at the $10^{-15}$ uncertainty level,'' Metrologia, vol. 43, pp. 109--120 (2006).
\bibitem{piester2008} D. Piester et al., ``Time transfer with nanosecond accuracy for the realization of International Atomic Time,'' Metrologia, vol. 45, pp. 185--198 (2008).
\bibitem{kim2008} J. Kim et al., ``Drift-free femtosecond timing synchronization of remote optical and microwave sources,'' Nat. Phot. vol. 2, pp. 733--736 (2008).
\bibitem{whiterabbit} P. Moreira et al., ``White Rabbit: Sub-nanosecond timing distribution over Ethernet,'' IEEE Symp. Prec. Clock Synchr. Meas. Contr. Comm., pp. 58--62 (2009).
\bibitem{rost2012} M. Rost et al., ``Time transfer through optical fibres over a distance of 73 km with an uncertainty
below 100 ps,'' Metrologia, vol. 49, pp. 772--778 (2012).
\bibitem{lopez2012time} O. Lopez et al., ``Simultaneous remote transfer of accurate timing and optical
frequency over a public fiber network,'' Appl. Phys. B, DOI 10.1007/s00340-012-5241-0 (2012).
\bibitem{wang2012} B. Wang et al., ``Precise and continuous time and frequency synchronisation at the $5\times10^{-19}$ accuracy level,''  Sci. Rep., vol. 2, pp. 1--5 (2012).
\bibitem{sliwczynski2013} \L{}. \'Sliwczy\'nski et al., ``Dissemination of time and RF frequency via a stabilized fibre optic link over a distance of 420 km,'' Metrologia, vol. 50, pp. 133--145 (2013).
\bibitem{levine2008} J. Levine, ``A review of time and frequency transfer methods,'' Metrologia, vol. 45, pp. S162--S174 (2008).
\bibitem{predehl2012} K. Predehl et al., ``A 920-kilometer optical fiber link for frequency metrology at the 19th decimal place,'' Science, vol. 336, pp. 441--444 (2012).
\bibitem{takamoto2011} M. Takamoto, T. Takano, and H. Katori, ``Frequency comparison of optical lattice clocks
beyond the Dick limit,'' Nat. Phot., vol. 5, pp. 288--292 (2011).
\bibitem{hinkley2013} N. Hinkley et al., ``An atomic clock with $10^{-18}$ instability,'' Science, pp. 1215--1218 (2013).
\bibitem{calosso2013} C. Calosso et al., ``Frequency transfer via a two-way optical phase comparison on a multiplexed fiber network,'' arXiv:1308.2377v2.
\bibitem{boumard2009} S. Boumard, and A. M\"ammel\"a, ``Robust and accurate frequency and timing synchronization using chirp signals,'' IEEE Trans. Broadcast., vol. 55, pp. 115--123 (2009). 
\bibitem{grosche2013}  G.  Grosche,  ``Eavesdropping  time  and  frequency:  phase noise  cancellation  along  a  time-varying  path,  such  as  an optical fiber,'' arXiv:1309.0728 [physics.optics] (2013). 
\bibitem{bercy2014} A. Bercy et al., ``In-line extraction of an ultra-stable frequency signal  over an optical fiber link,'' arXiv:1402.1307 [physics.optics] (2014).
\bibitem{djerroud2010} K. Djerroud et al., ``Coherent optical link through the turbulent atmosphere,'' Opt. Lett., vol. 35, pp. 1479--1481 (2010).
\bibitem{chiodo2013} N. Chiodo, K. Djerroud, O. Acef, A. Clairon, and P. Wolf, ``Lasers for coherent optical satellite links with large dynamics,'' Appl. Opt., vol. 52, pp. 7342--7351 (2013).
\bibitem{williams2008} P. A. Williams, W. C. Swann, and N. R. Newbury, ``High-stability transfer of an optical frequency over long fiber-optic links," J. Opt. Soc. Am. B, vol. 25, pp. 1284--1293 (2008).
\bibitem{lopez2012frequency} O. Lopez et al., ``Ultra-stable long distance optical frequency distribution using the Internet fiber network," Opt. Exp., vol. 20, pp. 23518--23526 (2012).
\bibitem{telle2002} H. R. Telle, B. Lipphardt and J. Stenger, "Kerr-lens, mode-locked lasers as transfer
oscillators for optical frequency measurements," Appl. Phys. B vol. 74, pp. 1--6 (2002).
\bibitem{lopez2010} O. Lopez et al., ``High-resolution microwave frequency dissemination on an 86-km urban optical link,'' Appl. Phys. B, vol. 98, pp. 723--727 (2010).
\bibitem{schediwy2012} S. Schediwy et al., ``Microwave frequency transfer with optical stabilisation,'' IEEE Proc. EFTF 2012, pp. 211--213 (2012).
\bibitem{rohde2013} F. Rohde, E. Benkler, and H. Telle, ``High contrast, low noise selection and amplification
of an individual optical frequency comb line,'' Opt. Lett. vol. 38, pp. 103--105 (2013) and refs. therein.
\bibitem{zou2013} X. Zou et al., ``All-fiber optical filter with an ultranarrow and rectangular spectral response,'' Opt. Lett., vol. 38, pp. 3096--3098 (2013).
\bibitem{fortier2011} T. M. Fortier et al., ``Generation of ultrastable microwaves via optical frequency division,'' Nat. Phot., vol. 5, pp. 425--429 (2011).
\bibitem{kramer2001} G. Kramer, and W. Klische, ``Multi-channel synchronous digital phase recorder,'' Proc. IEEE IFCS 2001, pp. 144--151 (2001).
\bibitem{dawkins2007} S. T. Dawkins, J. J. McFerran, and A. N. Luiten, ``Considerations on the Measurement of
the Stability of Oscillators with Frequency Counters,'' IEEE Trans. Ultras. Ferroel. Freq. Control vol. 54, pp. 918--925 (2007).
\bibitem{riley2008} W. Riley, ``Handbook of frequency stability analysis,'' NIST Special Publication 1065 (2008).
\bibitem{droste2013} S. Droste et al., ``Optical-frequency transfer over a single-span 1840 km fiber link,'' Phys. Rev. Lett. vol. 111, 110801 (2013). 
\bibitem{footnoteHarressSagnac} Harress before 1912 performed a rotational experiment \cite{harzer1914}, where the setup, aimed at investigations of the Lorentz/Fizeau drag coefficient, involved a rotating interferometer where the light travelled entirely in glass ($n=1.57$). At about the same time Laue predicted the rotational delay based on special relativity \cite{laue1911}.
\bibitem{laue1911} M. von Laue, ``{\"U}ber einen Versuch zur Optik bewegter K{\"o}rper,'' M{\"u}nch. Sitz.-Ber., 405--412 (1911) and refs. therein.
\bibitem{sagnac1913a} M. G. Sagnac, ``L'\'ether lumineux d\'emontr\'e par l'effet du vent relatif d'\'ether dans un interf\'erom\`etre en rotation uniforme,'' Compt. Rend. Acad. Sci. Paris, vol. 157, pp. 708--710 (1913).
\bibitem{sagnac1913b} M. G. Sagnac, ``Sur la preuve de la r\'ealit\'e de l'\'ether lumineux par l'exp\'erience de l'interf\'erographe tournant,'' Compt. Rend. Acad. Sci. Paris, vol. 157, pp. 1410--1413 (1913).
\bibitem{harzer1914} P. Harzer, ``{\"U}ber die Mitf{\"u}hrung des Lichtes in Glas und die Aberration,'' Astron. Nachr., vol. 198, pp. 377--392 (1914).
\bibitem{knopf1920} O. Knopf, ``Die Versuche von F. Harress {\"u}ber die Geschwindigkeit des Lichtes in bewegten K{\"o}rpern,'' Ann. d. Phys. vol. 62, pp. 389--447 (1920).
\bibitem{laue1919} M. von Laue, ``Zum Versuch von F. Harress,'' Ann. d. Phys. vol. 62, pp. 448--463 (1920).
\bibitem{post1967} E. J. Post, ``Sagnac effect,'' Rev. Mod. Phys vol. 39, pp. 475--493 (1967).
\bibitem{allan1985} D. W. Allan, M. A. Weiss, and N. Ashby, ``Around-the-world relativistic Sagnac experiment,'' Science vol. 228, pp. 69--70 (1985).
\bibitem{sotiropoulos2013} N. Sotiropoulos et al., ``Delivering 10 Gb/s optical data with picosecond timing uncertainty over 75 km distance," Opt. Expr. vol. 21, pp. 32643--32654 (2013).
\bibitem{wang2003} R. Wang et al., ``Modified Sagnac experiment for measuring travel-time difference between counter-propagating light beams in a uniformly moving fiber,'' Phys. Lett. A vol. 312, pp. 7--10 (2003).
\bibitem{arditty1981} H. J. Arditty, and H. C. Lef\`{e}vre, ``Sagnac effect in fiber gyroscopes,'' Opt. Lett. vol. 6, pp. 401--403 (1981).
\bibitem{itug652} International Telecommunication Union, ``Recommendation  ITU-T  G.652 11/2009'' (2009).




\end{thebibliography}
\end{document}